# Dual-Band Mid-Infrared Imaging for Measurement of Fuel Concentration and Temperature in a Supersonic Cavity

Loren C. Hahn[1]
*Kent State University, Kent, Ohio 44243*

Erik L. Braun[2] and Brendan J. McGann[2]
*Innovative Scientific Solutions, Inc., Dayton, Ohio 45459*

Katherine C. Opacich[3] and Timothy M. Ombrello[4]
*U.S. Air Force Research Laboratory, Wright-Patterson Air Force Base, Ohio, 45433, USA*

**A dual-band mid-infrared absorption imaging method is demonstrated for spatially resolved measurement of ethylene concentration and temperature in a Mach-2 cavity flow. The method used narrowband filters selected to exploit the differing in-band temperature sensitivities of ethylene absorption, together with a broadband infrared source and high-frame-rate mid-wave infrared camera. Temperature-dependent, filter-weighted absorption cross-sections were used to measure two-dimensional ethylene number density and temperature fields. Measurements were obtained for cavity, curtain, and discrete fuel-injection configurations, including cases with the leading edge of a quasi-steady shock near the cavity. Cavity equivalence ratio measurements agreed well with independent extractive measurements and reproduced trends previously observed using hyperspectral imaging. A complementary single-band implementation using a temperature-independent filter provided instantaneous ethylene concentration fields at 100 Hz and revealed intermittent reductions in cavity equivalence ratio associated with plume deflection by the shock train leading edge. The dual-band approach provides quantitative full-field measurements with substantially simpler spectral instrumentation than hyperspectral imaging while retaining flexibility for time-resolved flow measurements.**

## I. Introduction

Accurate measurements of hydrocarbon fuel concentration and temperature are essential for understanding fuel-air mixing, ignition, and combustion processes in high-speed propulsion systems. In scramjet combustors, the fuel distribution within the flowpath strongly influences ignition delay, flame stabilization, combustion efficiency, and overall engine performance [1]. Because these flows are highly compressible and turbulent, optical diagnostics have become indispensable experimental tools due to their non-intrusive nature and ability to provide spatially resolved measurements of important thermochemical quantities [2–4].

A wide variety of optical diagnostics have been developed for combustor measurements. Laser-based techniques, including tunable diode laser absorption spectroscopy, planar laser-induced fluorescence, and laser-induced breakdown spectroscopy provide quantitative measurements with excellent sensitivity but generally require complex optical systems or are limited to point or planar measurements [5–9]. Mid-infrared (mid-IR) imaging provides an attractive alternative because hydrocarbon fuels exhibit strong absorption and emission features in this spectral region, enabling quantitative, full-field measurements using comparatively simple instrumentation. Mid-IR diagnostics have been successfully applied to laboratory flames, industrial exhaust monitoring, and more recently to supersonic combustion systems [10–15].

[1] Assistant Professor, College of Aeronautics and Engineering; lhahn14@kent.edu
[2] Research Engineer
[3] Research Aerospace Engineer, Air Warfare Directorate.
[4] Senior Research Aerospace Engineer, Air Warfare Directorate.

Fuel mixing in nonreacting flows has been studied by measuring the absorption of infrared light by the fuel. Rhoby et al. [16] demonstrated quantitative hyperspectral mid-IR absorption imaging in the same Mach-2 model scramjet cavity investigated in the present work. By coupling a Fourier-transform spectrometer with a mid-IR camera, complete spectra were acquired at every pixel, enabling highly accurate reconstruction of two-dimensional temperature and hydrocarbon concentration fields. Although this approach provides excellent quantitative accuracy, the interferometric instrumentation is complex and expensive, and the spectral scanning required to reconstruct each image limits the acquisition rate for transient flows.

A simpler alternative uses narrowband spectral filters with a broadband infrared source to measure absorption at selected wavelengths. Ombrello et al. [17] demonstrated single-band absorption imaging to quantify hydrocarbon fuel distributions within a scramjet cavity by selecting a spectral region with minimal temperature dependence of the absorption cross-section. More recently, van der Lee et al. [18] employed a similar technique to investigate liquid-fuel mixing and entrainment in a cavity-based scramjet combustor. While this approach significantly simplifies the experimental system, assuming a temperature-independent absorption cross-section can introduce errors when substantial temperature variations exist within the flow.

Dual-wavelength absorption techniques overcome this limitation by exploiting the different temperature sensitivities of two absorption bands to simultaneously determine temperature and fuel concentration. Klingbeil et al. [19] and Pyun et al. [20] demonstrated a similar concept using a two-wavelength mid-IR laser absorption sensor capable of quantitative and time-resolved line-of-sight measurements. With the proven success of using multiple wavelength bands to extract single line-of-sight information, there is motivation to explore 2-D mid-IR measurements using imaging-based techniques.

The present work extends the dual-wavelength concept to full-field infrared imaging using a multispectral mid-IR camera equipped with interchangeable narrowband filters. Two carefully selected spectral bands within the ethylene absorption spectrum are used to simultaneously determine the path-integrated ethylene concentration and temperature from two-dimensional absorption images. Measurements are obtained in the same Mach-2 model scramjet cavity investigated by Rhoby et al. [16] for a range of fuel injector configurations in a nonreacting flow. Compared with hyperspectral imaging, the proposed approach achieves comparable quantitative capability using substantially simpler and lower-cost instrumentation while eliminating spectral scanning, thereby enabling instantaneous image acquisition. Compared with laser-based dual-wavelength absorption, the method provides spatially resolved measurements over the entire field of view. The resulting measurements reveal the fuel concentration and temperature distributions produced by a wide range of injector configurations and are used to investigate the influence of an upstream shock train on cavity fueling. The flexibility of the imaging approach is also demonstrated through single-filter measurements capable of capturing unsteady flow phenomena. Together, these measurements provide new insight into the local fuel and temperature fields that govern ignition and flameholding in cavity-based scramjet combustors.

## II. Experimental Setup

### a. Tunnel Configuration

Experiments were performed in the Research Cell 19 supersonic wind tunnel at Wright-Patterson Air Force Base [21]. The flowpath has a constant-area rectangular isolator section with a width of 152.4 mm, a height of 50.8 mm, and a length of 177.8 mm. The isolator is followed by a combustor section with a bottom wall that diverges at $2.5^{\circ}$. The cavity was within the bottom wall of the combustor with a depth of 16.5 mm, length-to-depth ratio of 4 (length defined from the step to the middle of the closeout ramp), and a $22.5^{\circ}$ ramped closeout. Gaseous ethylene was injected using three separate configurations. The first configuration was direct cavity fueling, where a single plenum feeds 11 upstream-facing 1.98 mm diameter holes within the cavity closeout ramp. The holes are equally spaced across the entire span of the flowpath. The second configuration was discrete upstream fueling. Three 3.18 mm injectors were positioned 28.6 mm upstream of the cavity step. The discrete upstream injectors were spaced 38.1 mm apart with the middle of the three on the streamwise centerline of the tunnel. Thirdly, upstream curtain fueling was used. Curtain fueling used 21 injectors (11 of 1.3 mm and 10 of 0.5 mm alternating and equally-spaced) located 55.9 mm upstream of the cavity step. Figure 1 contains a diagram of the cavity geometry including the injector configurations.

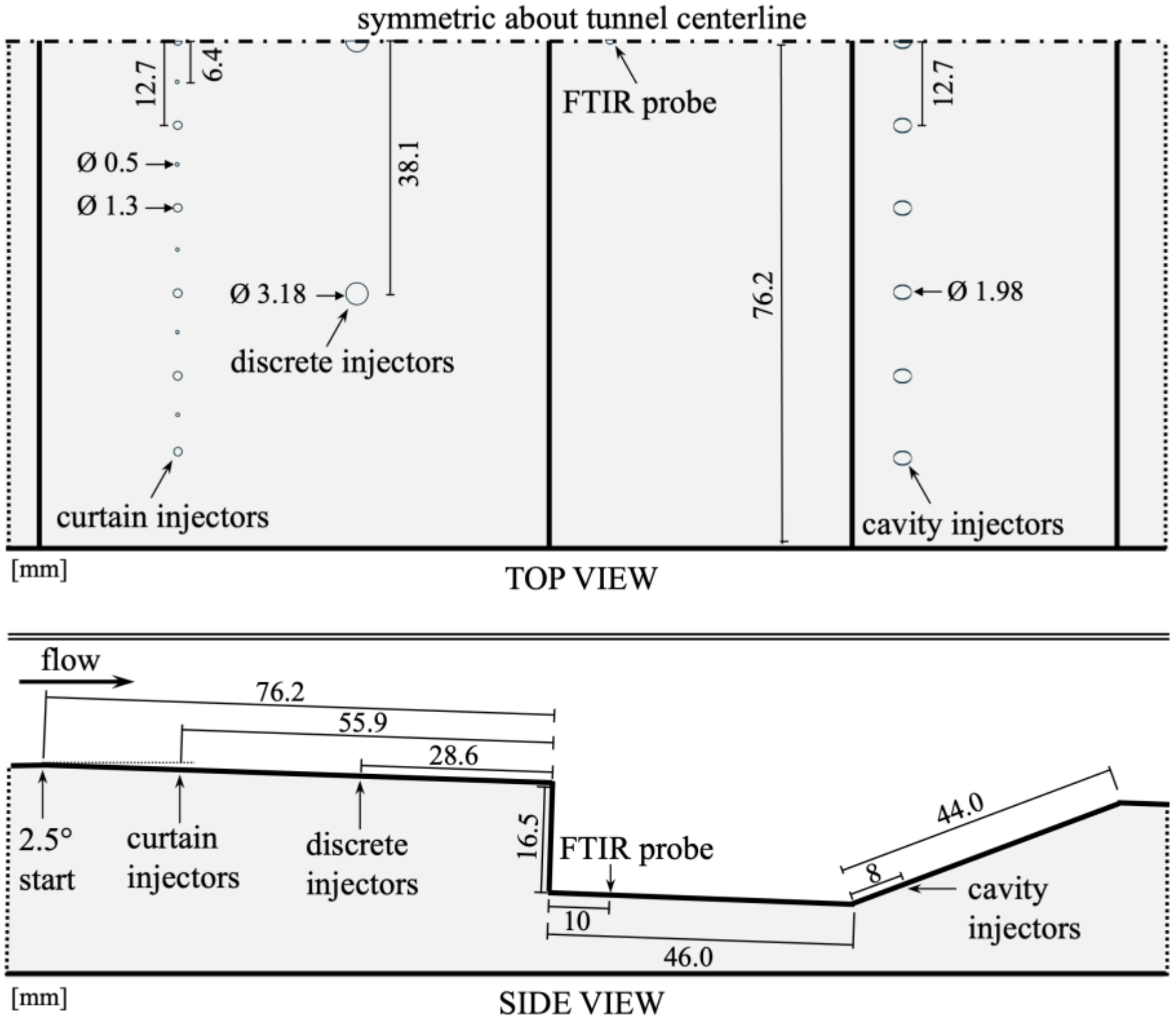

Figure 1: Experimental cavity diagram (modified based on [16]).

For all experimental runs, Mach number, $M$, exiting the nozzle was 2, the total temperature, $T_o$, was 589 K, and the total pressure, $P_o$, was 689.5 kPa. Each experimental point consisted of a specific injector configuration and a set fuel flow rate controlled by an upstream sonic nozzle. There were five experimental configurations where the effect of varied fuel flow rate was explored. These configurations and the flow rates considered are provided in Table 1. The fuel flow rates were standardized to 294 K and 101.3 kPa. For the cavity and curtain injector configurations, fuel was supplied to a plenum upstream of the injector port arrays, resulting in each array being supplied with a fixed pressure and therefore a constant flow rate for a given port diameter. The three ports in the discrete injector configuration were fueled independently. There were cases where all three ports were fueled as well as cases where only a single (center) port was fueled. Finally, there were cases using three discrete injectors in which the tunnel was backpressured using downstream mechanical valves. The backpressure cases formed a shock train (ST) which was positioned such that the shock train leading edge (STLE) was located near the cavity step. This position was chosen because the STLE exhibited significant unsteadiness and was a useful case to demonstrate the utility of instantaneous imaging with a single filter. The runs with the ST present are denoted “3 discrete + ST”. Figure 2 shows the tunnel pressure distribution for the baseline cases and the backpressured case.

Table 1: Experimental fueling configurations and flow rates.

| Configuration | Ethylene Flow Rate [slpm] |
|---|---|
| cavity | 50, 100, 150, 200, 250, 300 |
| curtain | 250, 500, 1000, 1500, 2000 |
| center discrete | 250, 500, 1000 |
| 3 discrete | 100, 250, 500, 1000, 1500, 2000 |
| 3 discrete + ST | 500, 2000 |

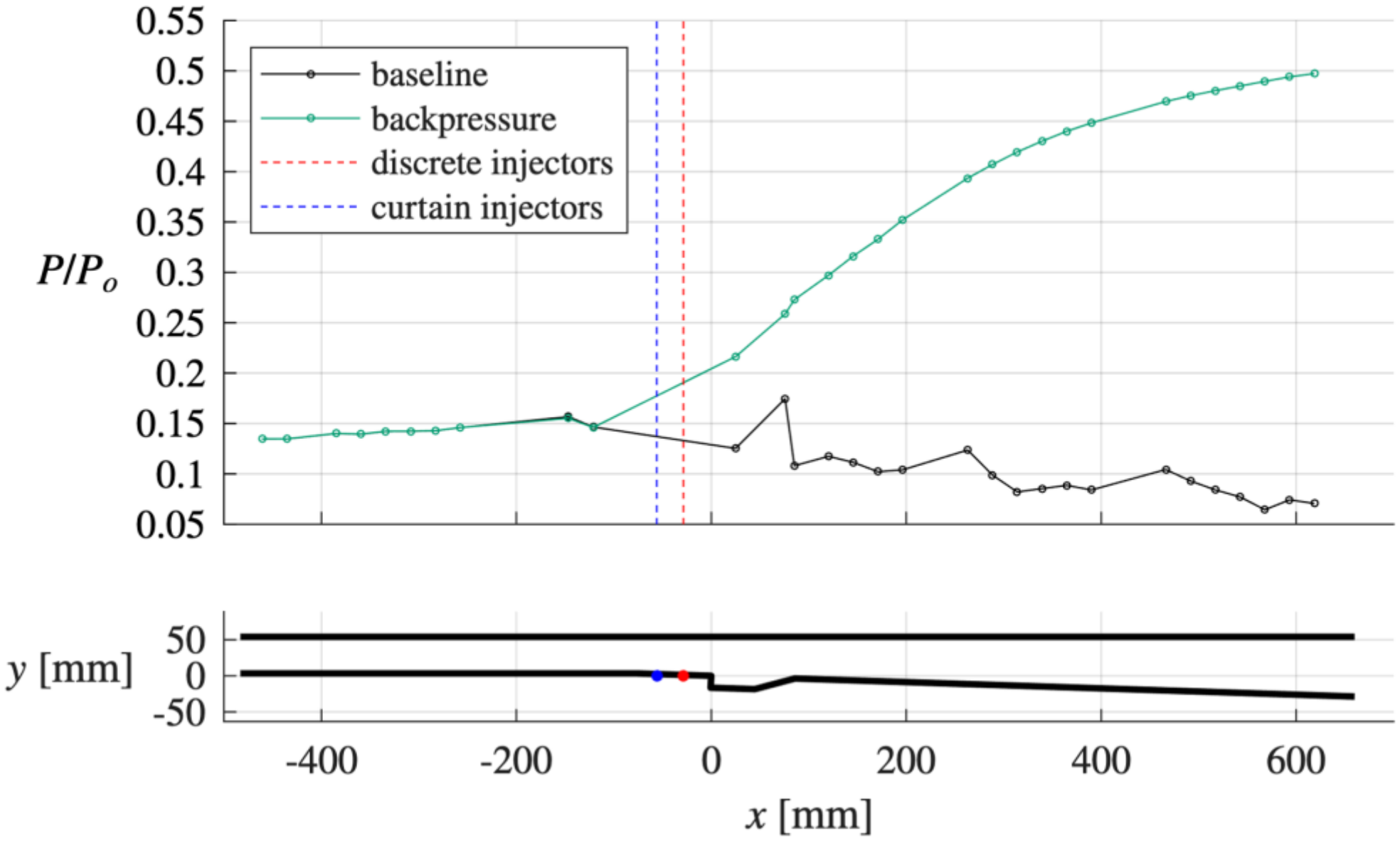

Figure 2: Tunnel pressure distributions (top) aligned with flowpath wall dimensions (bottom).

### b. Diagnostic Configuration

Two UV grade fused silica windows (Corning 7980) were installed on the tunnel sidewalls to provide optical access to the test section. The Corning 7980 window provided sufficiently high transmission of the IR bands investigated in this work to allow for strong absorption signals. On one side of the test section, a CI Systems SR-800N-8HT extended-area blackbody was positioned normal to the window and maintained a constant temperature of 873 K throughout the experiment. Images of the test section (focused at the center plane) were captured using a Telops MS-M3k mid-wave infrared camera. The camera captures light using a sensor array of 320x256 pixels which corresponded to a spatial sampling of 375 μm/pixel at the focal plane when using a Stingray 50mm f/2.5 lens. The MS-M3k is equipped with a rotating 8-position filter wheel which was operated at its maximum revolution rate which resulted in images being captured sequentially at an overall rate of 800 Hz and a filter-specific rate of 100 Hz. The filters were chosen to target bands of strong emission and absorption by hydrocarbons. The filters were manufactured by Spectrogon and their details are provided in Table 2. The second and third elements of the part number describe the filter center wavelength and full-width at half-maximum (FWHM) in nm.

Table 2: Narrowband filters used for multispectral imaging.

| Filter Wheel Position | Part number |
|---|---|
| 1 | NB-3331-020 |
| 2 | NB-3365-020 |
| 3 | NB-3385-025 |
| 4 | NB-3375-025 |
| 5 | NB-3421-025 |
| 6 | NB-3395-025 |
| 7 | NB-3465-020 |
| 8 | NB-3405-025 |

The present work investigated absorption by ethylene and utilized the experimental data from Alrefae et al. [22]. These data contain the spectra of absorption cross-section, $\varepsilon$, of various hydrocarbon fuels at atmospheric pressure across varied temperatures. Figure 3 contains a band of the absorption cross-section spectra from Alrefae et al. [22] with the present experiment filter center wavelengths (CWLs) overlaid. The spectra were smoothed with a moving average filter of width 5 nm for clarity. It was observed that the absorption cross-sections showed varied dependencies on the temperature across the mid-IR spectrum. This temperature dependency is the basis of the diagnostic technique outlined in the following section.

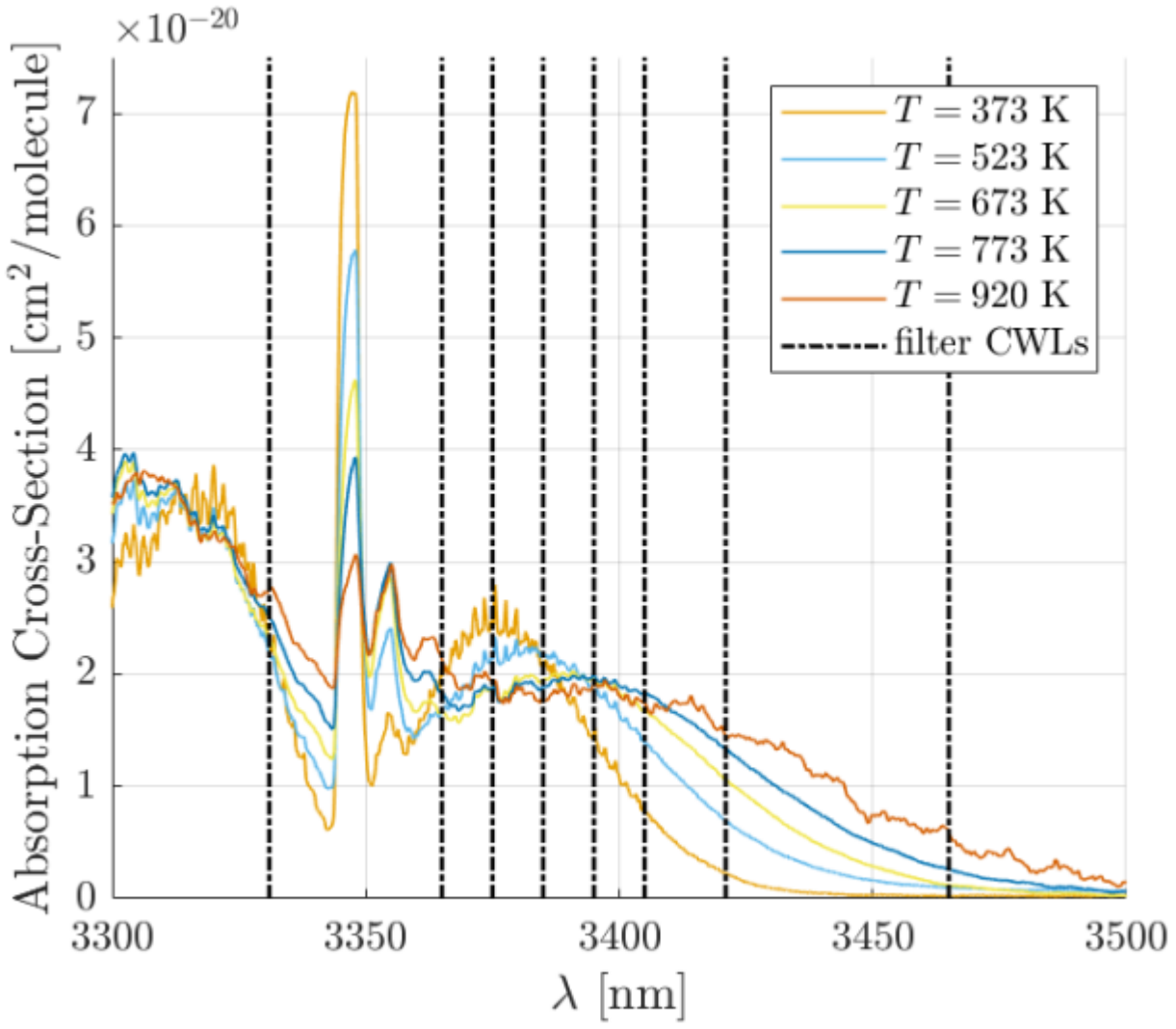

Figure 3: Ethylene absorption cross-section spectra at varied temperature [22] with current experimental filter center wavelengths (from Table 2) indicated.

The MS-M3k acquired data for 5 seconds for every image set, corresponding to 500 images for each of the 8 filters. Every image was captured with an exposure time of 10 µs. An NLIR MIDWAVE Spectrometer equipped with a collimator was used to capture the mid-IR spectrum during fueling. The spectrometer collected light over a range of 2.0-5.0 µm along a line-of-sight column in space. The collimating optic was positioned normal to the test section optical windows and aligned with the center of the cavity. The spectrum was acquired by averaging over 50 measurements captured with an exposure time of 40 ms. A Fourier-transform infrared spectrometer (FTIR) was used to extractively sample and measure the ethylene mole fraction along the cavity floor and ultimately served to validate the dual-band absorption method. The same FTIR system outlined by Rhoby et al. [16] and Ombrello et al. [17] was used in the present work.

## III. Diagnostic Techniques

### a. Dual-Band Mid-IR Absorption Method

The Beer-Lambert law at a given wavelength relates the transmitted intensity, $I$, through an absorbing medium to the incident intensity, $I_i$, according to

$$I = I_i e^{-\varepsilon N L}, \tag{1}$$

where $\varepsilon$ is the absorption cross-section, $N$ is the molecular number density, and $L$ is the optical path length through the absorbing medium. Because the measured mid-IR signal also contains thermal emission from the tunnel windows, four acquisitions were used to isolate the attenuation of the blackbody radiation by the fuel.

For each fueling condition, four separate acquisitions were obtained. The first acquisition measured the blackbody signal through the tunnel without fueling and is denoted $I_0$. The second acquisition was obtained without fueling but with an insulated shield positioned between the blackbody and the tunnel, thereby removing the contribution from the blackbody; this signal is denoted $I_{0,s}$. The third and fourth acquisitions repeated these measurements with fuel injection and are denoted $I_1$ and $I_{1,s}$, respectively.

Based on the mid-IR light path shown in Figure 4, the measured intensity for each acquisition can be expressed as follows:

$$I_0 = I_B \tau_{w1} \tau_{w2} + I_{w1} \tau_{w2} + I_{w2}, \tag{2}$$

$$I_{0,s} = I_{w1} \tau_{w2} + I_{w2},$$

$$I_1 = I_B \tau_{w1} e^{-\varepsilon NL} \tau_{w2} + I_{w1} e^{-\varepsilon NL} \tau_{w2} + I_{w2},$$

$$I_{1,s} = I_{w1} e^{-\varepsilon NL} \tau_{w2} + I_{w2},$$

where $I_B$ is the light emitted from the blackbody, $I_{w1}$ and $I_{w2}$ are the thermal emission contributions from windows 1 and 2, respectively, $\tau_{w1}$ and $\tau_{w2}$ are their respective transmittances. Subtracting the corresponding shielded acquisition from each unshielded acquisition removes the window-emission contributions. The ratio of the fueled and unfueled signals is therefore

$$X = \frac{I_1 - I_{1,s}}{I_0 - I_{0,s}} = e^{-\varepsilon NL}. \tag{3}$$

The molecular number density can then be determined from

$$N = \frac{-\ln(X)}{L\varepsilon}. \tag{4}$$

The absorption cross-section data were available at five discrete temperatures. To permit evaluation at intermediate temperatures, the temperature dependence of the absorption cross-section was represented independently at each wavelength. At each wavelength, a quadratic polynomial was fit to the five measured values of $\varepsilon$ as a function of temperature, yielding a continuous function $\varepsilon(T)$ at that wavelength. Repeating this procedure across the measured spectrum produced a two-dimensional representation of the absorption cross-section, $\varepsilon(\lambda, T)$. Because each camera measurement is integrated over the finite spectral bandwidth of a narrowband filter, a filter-weighted absorption cross-section was calculated as

$$\varepsilon_f(T) = \frac{\int \tau_f(\lambda) \varepsilon(\lambda, T) d\lambda}{\int \tau_f(\lambda) d\lambda}, \tag{5}$$

where $\tau_f(\lambda)$ is the filter transmission curve. Measurements obtained using two filters, denoted $a$ and $b$, provide two independent relations for the same molecular number density

$$N = \frac{-\ln(X_a)}{L\varepsilon_a(T)} = \frac{-\ln(X_b)}{L\varepsilon_b(T)}. \tag{6}$$

Equation (6) is solved for $T$, after which $N$ is obtained from either absorption band. The two filters are selected such that the ratio $\varepsilon_a(T)/\varepsilon_b(T)$ varies monotonically over the temperature range of interest, proving a unique temperature solution.

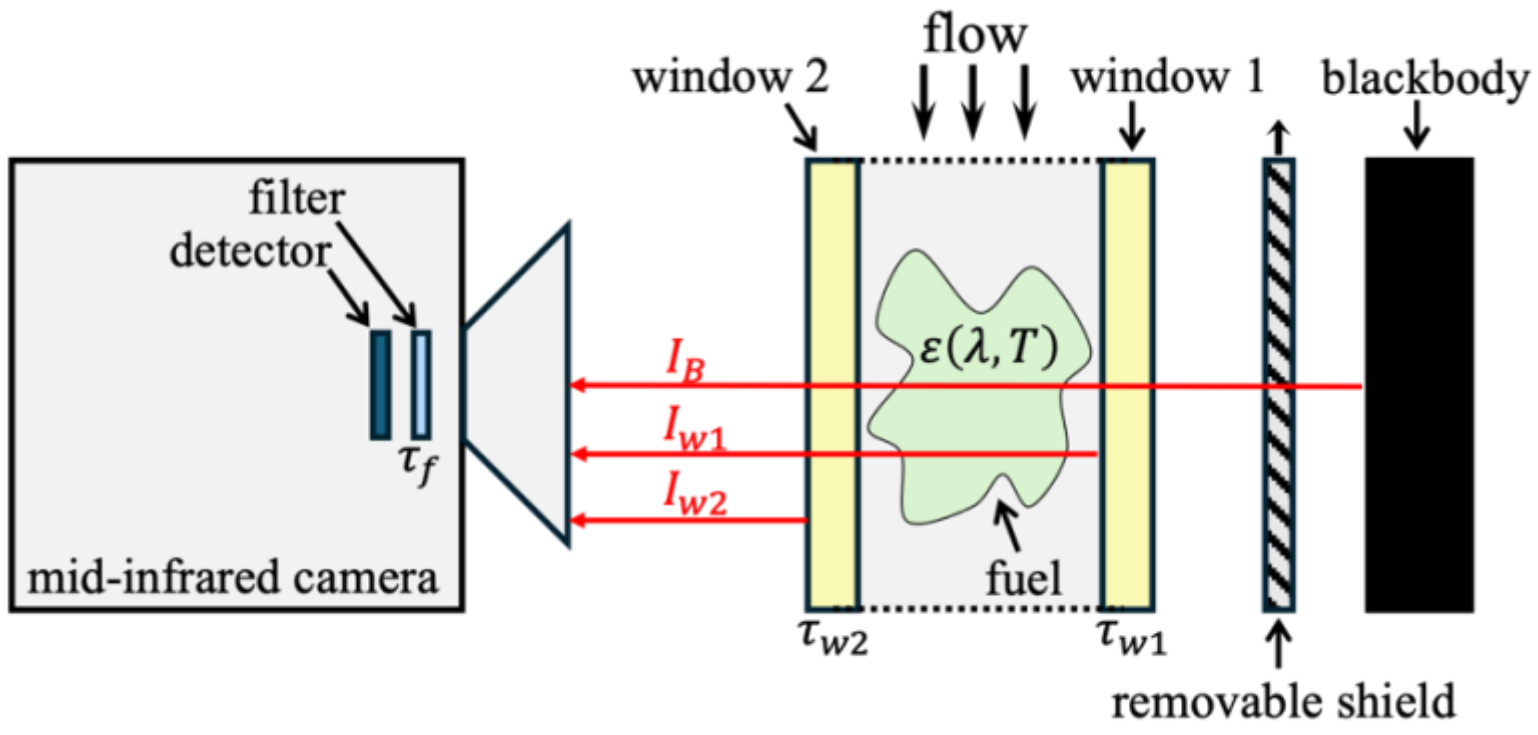

Figure 4: Mid-infrared light path for the dual-band absorption method.

## b. Uncertainty Quantification

Two sources of uncertainty were considered: uncertainty associated with the absorption cross-section fits and experimental variability in the measured images. For each filter, the uncertainty associated with the continuous representation of the absorption cross-section was estimated from the agreement between the quadratic fit and the five discrete temperature-dependent values. The filter-specific fitting uncertainty, $u_{\varepsilon_f}$, was defined as the root mean square error (RMSE),

$$u_{\varepsilon_f} = \sqrt{\frac{1}{n}\sum_{i=1}^{n}(\varepsilon_{f,\text{fit}}(T_i) - \varepsilon_f(T_i))^2}, \quad n = 5, \tag{7}$$

where $\varepsilon_{f,\text{fit}}(T_i)$ is the value predicted from the continuous representation and $\varepsilon_f(T_i)$ is the value calculated from the measured spectrum at temperature $T_i$. For a given filter pair, Eq. (6) can be expressed in terms of the ratio of the absorption cross-sections $R_\varepsilon(T) = \frac{\varepsilon_a(T)}{\varepsilon_b(T)} = \frac{\ln(X_a)}{\ln(X_b)}$. Assuming the uncertainties of the filter-weighted absorption cross-sections are independent, a linear propagation through the ratio yields

$$u_{T,f} = \frac{u_{R_\varepsilon}}{|dR_\varepsilon/dT|} = \frac{\sqrt{(u_{\varepsilon_a}/\varepsilon_a)^2 + (u_{\varepsilon_b}/\varepsilon_b)^2}}{|\alpha_a - \alpha_b|}. \tag{8}$$

where the relative temperature sensitivities are defined as $\alpha_f = \frac{1}{\varepsilon_f}\frac{d\varepsilon_f}{dT}$ for brevity. The above expression demonstrates that temperature sensitivity is governed by the difference between the relative temperature sensitivities of the two absorption bands. Filter pairs for which these quantities are similar provide poor temperature sensitivity and correspondingly large uncertainties.

Experimental variability was evaluated from spatial fluctuations within the measured temperature fields. Smooth functions were fit independently along the horizontal and vertical directions within the cavity, yielding $T_{\text{fit},x}$ and $T_{\text{fit},y}$.

The RMSE between the measured temperature values at each pixel and these fitted distributions was calculated in each direction, and the resulting experimental uncertainty was defined

$$u_{T,r} = \sqrt{RMSE(T(x))^2 + RMSE(T(y))^2} \,. \tag{9}$$

The total temperature uncertainty is then

$$u_{T,t} = \sqrt{u_{T,f}^2 + u_{T,r}^2} \,. \tag{10}$$

The uncertainty of the molecular number density associated with the absorption cross-section is calculated as

$$u_{N,f} = N\sqrt{\left[\frac{\alpha_b}{\alpha_a - \alpha_b}\frac{u_{\varepsilon_a}}{\varepsilon_a}\right]^2 + \left[\frac{\alpha_a}{\alpha_a - \alpha_b}\frac{u_{\varepsilon_b}}{\varepsilon_b}\right]^2} \,. \tag{11}$$

Analogous to the temperature calculation, spatial variability in the measured number density field was quantified as

$$u_{N,r} = \sqrt{RMSE(N(x))^2 + RMSE(N(y))^2} \,. \tag{12}$$

Finally, the total uncertainty of the molecular number density is

$$u_{N,t} = \sqrt{u_{N,f}^2 + u_{N,r}^2}. \tag{13}$$

The top panel of Fig. 5 shows the filter-weighted absorption cross-sections for three representative filters. Filter 3365 exhibits a decreasing absorption cross-section with increasing temperature, whereas filter 3421 exhibits an increasing absorption cross-section with increasing temperature. In contrast, filter 3385 is comparatively insensitive to temperature over much of the range. The corresponding temperature uncertainties calculated from Eq. (8) are shown in the bottom panel.

The 3365-3421 pair maintains a relatively low uncertainty throughout the temperature range because the two filters have opposing relative temperature sensitivities. The 3385-3421 pair is similar, exhibiting a slightly elevated uncertainty over the entire range due to its decrease in the temperature sensitivity. The uncertainty of the 3365-3385 pair increases strongly at higher temperatures as their relative sensitivities approach one another, reducing the temperature dependence of $R_\varepsilon$. All pairwise combinations of the eight filters were evaluated using this procedure and the ten combinations having the lowest maximum value of $u_{T,f}$ over $200 < T < 600$ K are reported in Table 3. This range was chosen to bound the expected possible temperature range of the fuel in the experiments, which would be less than or equal to the stagnation temperature (589 K).

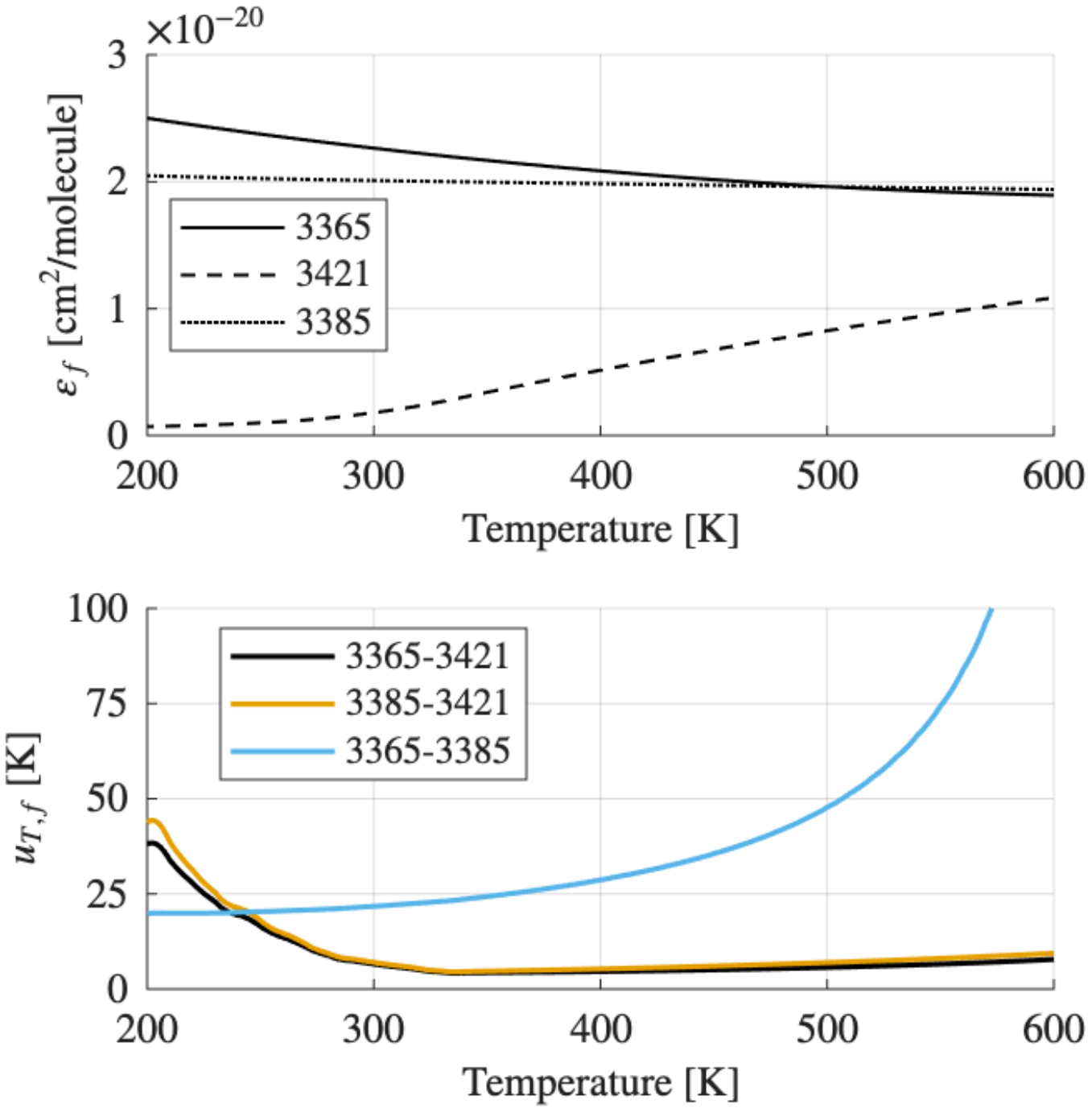

Figure 5: Transmission-weighted absorption cross-section curves for three representative filters (top) and associated temperature uncertainties from their combinations (bottom).

Table 3: Ranking of filter pairs based on lowest maximum $u_{T,f}$ over range 200 K < $T$ < 600 K. Pairs with higher maximum $u_{T,f}$ values are omitted.

| Rank | Pair | Maximum $u_{T,f}$ [K] | Minimum $u_{T,f}$ [K] |
|---|---|---|---|
| 1 | 3365-3421 | 38 | 4.2 |
| 2 | 3375-3421 | 40 | 4.2 |
| 3 | 3331-3421 | 42 | 4.5 |
| 4 | 3385-3421 | 44 | 4.5 |
| 5 | 3421-3395 | 47 | 4.6 |
| 6 | 3421-3405 | 49 | 4.7 |
| 7 | 3331-3375 | 72 | 63 |
| 8 | 3331-3365 | 73 | 39 |
| 9 | 3375-3405 | 81 | 19 |
| 10 | 3365-3405 | 83 | 16 |

## IV. Results

The dual-band method was applied using the 3365 and 3421 nm filters, which provided the lowest predicted temperature uncertainty over the range considered in Section III.b. The following sections first examine the measured absorption signals and then present the resulting ethylene concentration and temperature fields for the different fueling configurations. Comparisons with FTIR and previous hyperspectral measurements are used to evaluate the measurements. Finally, the approximately temperature-independent 3385 nm band (see Figure 5) is used independently to obtain instantaneous ethylene concentration fields and examine unsteady fuel redistribution by a shock train.

### a. Absorption Signal

Figure 6 shows the relative transmitted spectral intensity measured by the mid-IR spectrometer for the cavity fueling cases. Attenuation varied substantially across the 3000 to 3500 nm range as a result of the wavelength-dependent ethylene absorption cross-section. Increasing fuel flow generally produced greater attenuation, with the largest differences occurring near the strongest ethylene absorption features

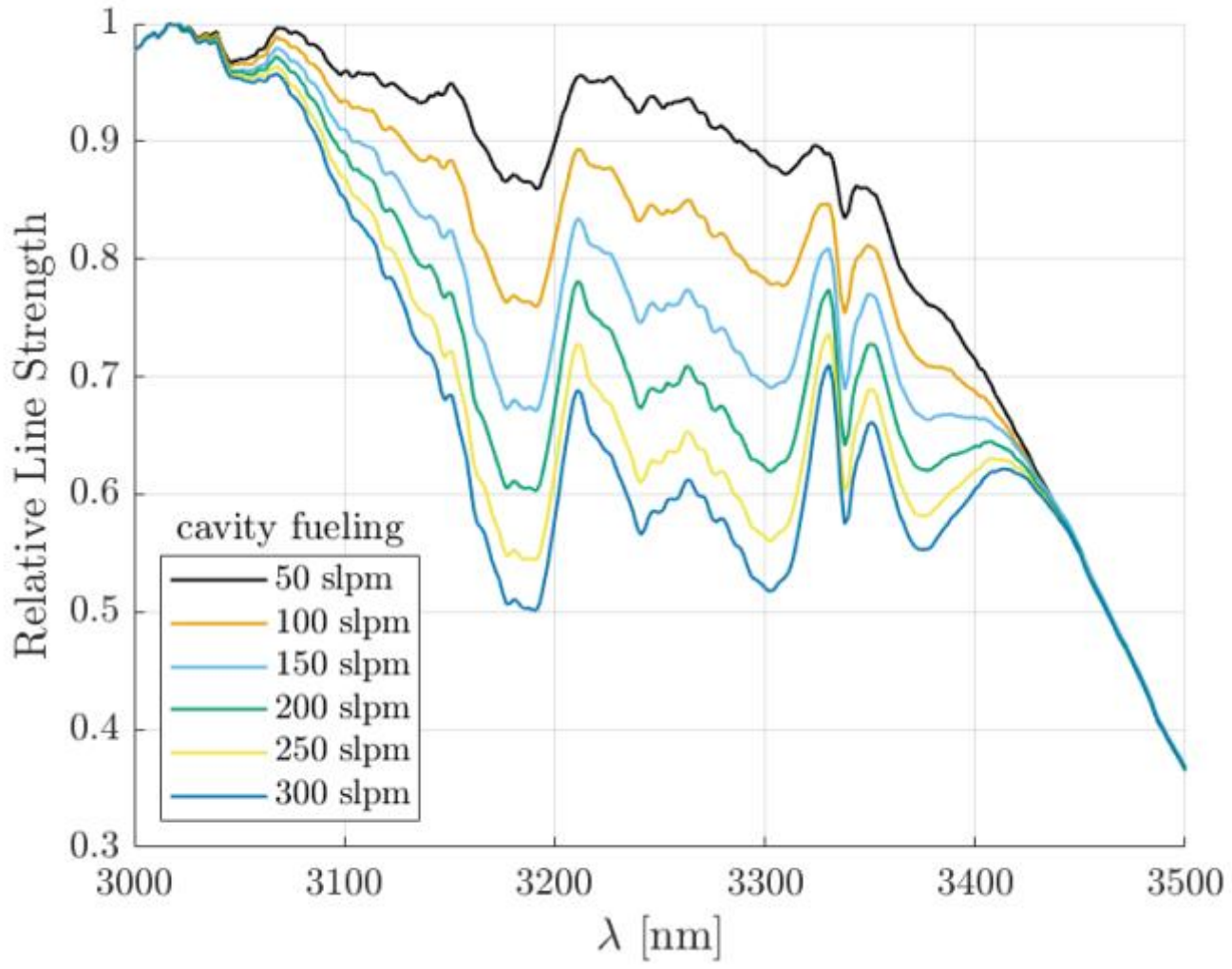

Figure 6: Relative transmitted spectral intensity measured by the mid-IR spectrometer for the cavity fueling cases.

Figure 7 shows the signal ratios $X_a$ and $X_b$ obtained using the 3365 and 3421 nm filters, respectively, for cavity fueling at 300 slpm. Both ratios approach unity in the freestream, as expected, and decrease within the cavity where ethylene is present. The magnitude of attenuation differs between the two filters, with $X_a$ lower than $X_b$ over much of the cavity. These differences result from the distinct absorption cross-sections and temperature sensitivities of the two bands and provide the information required to simultaneously determine ethylene concentration and temperature.

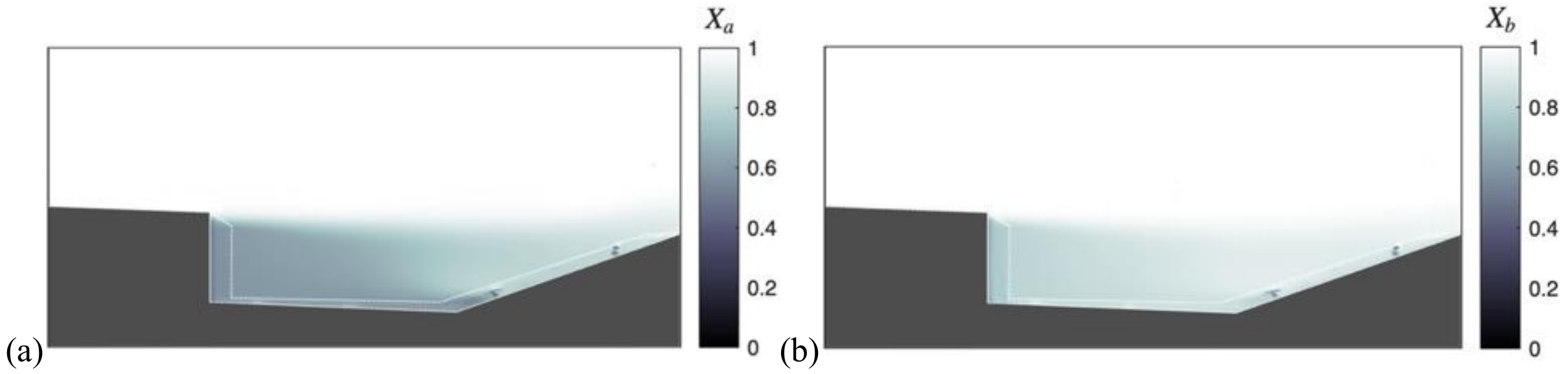

Figure 7: Signal ratios for cavity fueling at 300 slpm using the (a) 3365 nm and (b) 3421 nm filters.

### b. Fuel Concentration and Temperature Fields

The signal ratios $X_a$ and $X_b$ were obtained for each configuration listed in Table 1, and the dual-band inversion described in Section III.a was used to determine two-dimensional ethylene molar concentration and temperature fields.

Figure 8 shows the highest fuel rate case investigated for each configuration. Figure 8a presents the ethylene concentration and Fig. 8b presents the corresponding temperature.

For cavity fueling at 300 slpm, a relatively high ethylene concentration was observed along the cavity floor and near the downstream face of the step. The fuel remained underneath the cavity shear layer and the concentration decreased toward the cavity ramp. The average temperature varied little through much of the cavity, with a localized lower-temperature region near the cavity injectors that is closer to the room temperature value at which the fuel was delivered to the cavity plenum. A higher-temperature region was observed near the ramp, where the cavity shear layer impinges and forms a compression wave. These features are consistent with the measurements of Rhoby et al. [16].

Curtain fueling produced a comparatively uniform layer of fuel entering the measurement region upstream of the cavity. The ethylene concentration within this layer decreased with streamwise distance as the injected fuel mixed with the surrounding flow. Within the cavity, the concentration was comparatively uniform, while an increase was observed near the ramp where the flow was compressed and decelerated. The temperature upstream of the step was below the stagnation temperature but substantially above the freestream static temperature, consistent with fuel that had mixed with the incoming boundary-layer flow. The temperature decreased as the flow expanded around the step corner. Within the cavity, the measured temperature was relatively uniform near 500 K. The correspondence between the measured fuel temperature and the expected thermal properties of the primary flow suggests that spatially distributed fuel seeding may also provide a means of using the dual-band technique to interrogate flow temperature.

The center discrete injector at 1000 slpm produced a highly underexpanded jet that penetrated well into the primary flow. Ethylene concentration was highest near the injector and decreased downstream as the jet spread and accelerated. Very little fuel entered the cavity for this condition, and the resulting weak absorption prevented sufficient signal measurement over a substantial portion of the cavity. Near the injector, resolved temperatures approached 300 K, consistent with strong cooling during expansion of the fuel jet.

With three-discrete injectors at 2000 slpm, the fuel also penetrated substantially into the primary flow and became progressively leaner downstream. Compared with curtain fueling, the concentrated portion of the fuel plume penetrated farther from the wall and interacted less strongly with the cavity ramp. The cavity concentration remained relatively uniform. The temperature field exhibited a local minimum near the injectors and a distinct increase across the cavity shear layer as the injected fuel mixed with the warmer primary flow.

The effect of the shock train is shown by comparing the three-discrete-injector cases with and without backpressure. In the presence of the shock train, the fuel plume was deflected away from the cavity and farther into the core flow. This behavior is consistent with the separated and strongly deflected flow associated with a shock train leading edge [23]. Despite the upward deflection, the ethylene molar concentration within the cavity increased relative to the corresponding case without a shock train at the same fuel flow rate. The increase is consistent with the higher static density produced by compression through the shock train. The temperature field likewise increased over much of the measurement region, including near the cavity shear layer and step corner.

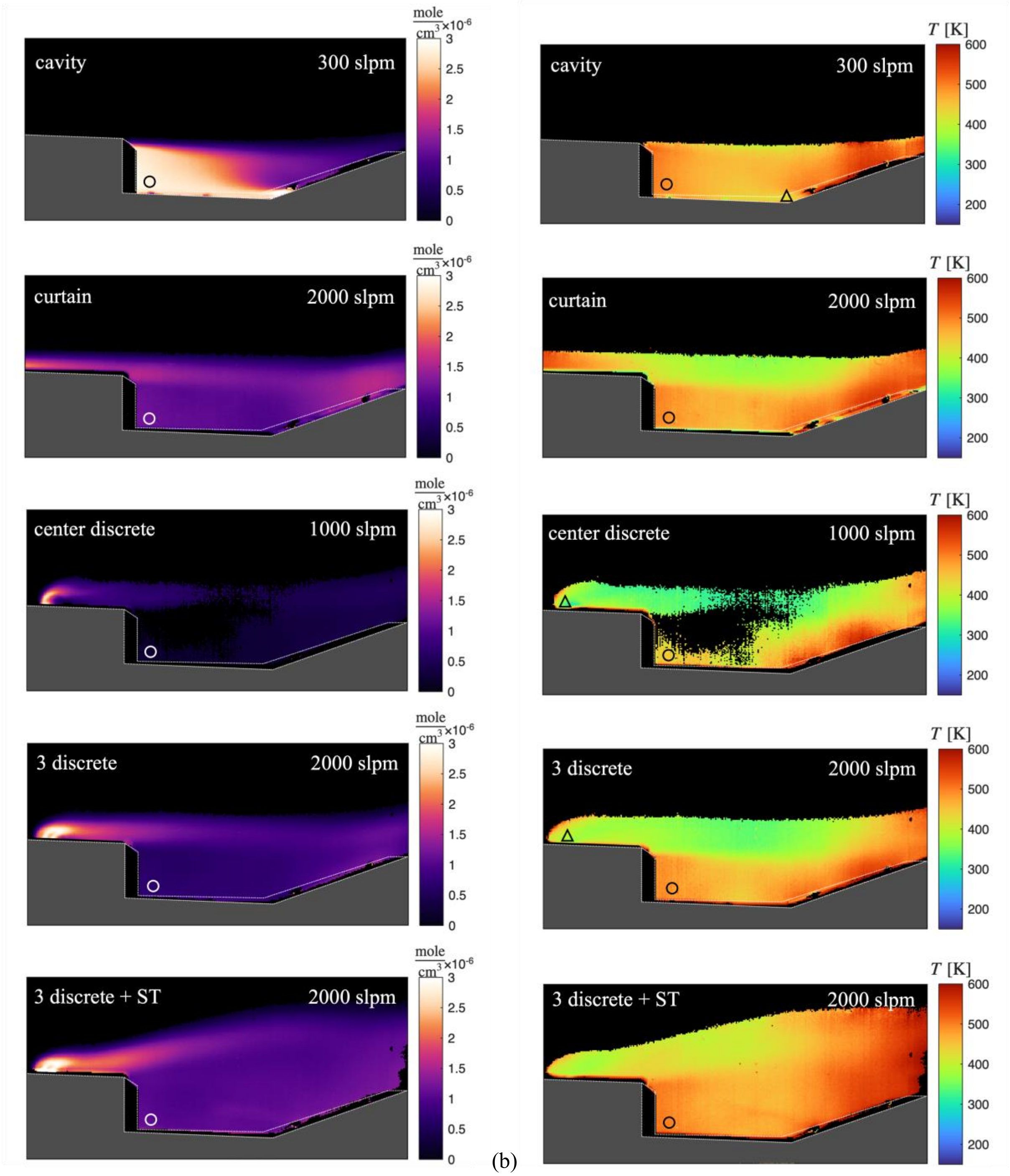

Figure 8: Two-dimensional dual-band measurements for the highest investigated fuel flow rate of each configuration: (a) ethylene molar concentration and (b) temperature. Circle markers indicate cavity probe locations and triangle markers indicate injector-region probe locations used in the subsequent analysis.

Quantitative values were extracted at the locations indicated in Fig. 8. Each value represents the average over a 3x3 pixel region. The cavity measurements were converted to fuel-air equivalence ratio, $\phi$, to facilitate comparison with the FTIR measurements and the previous hyperspectral measurements of Rhoby et al. [16]. Ethylene molar concentration was obtained from the dual-band measurement, while the air molar concentration was calculated using the measured static pressure at the cavity transducer. The pressure was assumed to be spatially uniform within the subsonic cavity recirculation.

Figure 9a compares the present cavity fueling measurements with the FTIR and hyperspectral measurements of Rhoby et al. [16]. The fuel flow rate was corrected for the difference in stagnation pressure between the two experiments. All four datasets show an approximately linear increase in cavity equivalence ratio with fuel flow rate. The FTIR measurements from the two experiments agree closely, indicating good repeatability between the separate test campaigns. The present dual-band measurements also agree closely with the FTIR measurements at lower fuel flow rates, with a modest deviation appearing at the highest flow rates. The hyperspectral measurements of Rhoby et al. [16] similarly produced slightly elevated equivalence ratios compared to the FTIR measurements, with the difference increasing at higher fuel flow rates.

Figure 9b compares the dual-band and FTIR measurements for the remaining injector configurations. The two diagnostics captured the same overall trends, although the FTIR values were generally slightly higher, particularly at the largest fuel flow rates. Curtain fueling produced an approximately linear increase in cavity equivalence ratio because additional fuel introduced into the upstream boundary layer resulted in a corresponding increase in fuel entrained into the cavity. The three-discrete-injector configuration behaved differently. The cavity equivalence ratio initially increased with fuel flow rate, reached a local maximum near 500 slpm, and then decreased slightly before approaching a plateau. Increasing the discrete-jet flow rate therefore increased both the amount of injected fuel and the penetration of the jet into the core flow. At sufficiently high flow rates, additional fuel primarily penetrated deeper into the core flow above the cavity shear layer rather than increasing the amount entrained into the cavity. A similar behavior was reported by Rhoby et al. [16].

The center discrete injector exhibited a local minimum in cavity equivalence ratio at 500 slpm. Comparison with the three-injector cases at matched flow rate per injector indicates that this condition corresponds approximately to the transition where increased jet penetration begins to reduce fuel delivery to the cavity. At still higher flow rates, the greater total fuel supply partially offsets this effect. The ST cases provide an additional contrast. For the same fuel flow rate as the non-ST case, the presence of the ST reduced the local cavity equivalence ratio, consistent with upward deflection of the fuel plume by the STLE. This is an important finding for the operation of a scramjet engine because the formation of a ST can lead to a rapid decrease in the cavity equivalence ratio. Abrupt changes in local equivalence ratio could lead to flameout and must be considered when operating at the threshold between scram and ram mode.

Several effects may contribute to the differences between the dual-band and FTIR measurements. First, the dual-band analysis uses time-averaged signal ratios, while the Beer-Lambert relation is nonlinear, such that temporal fluctuations are not retained by the averaged measurement. Second, the FTIR sampled directly from the cavity wall, whereas the imaging geometry prevented reliable near-wall measurement and were instead sampled at an offset of 3 mm above the wall. The extractive wall measurements may have therefore sampled a potentially locally richer region. Finally, the discrete injector configurations are nonuniform in the spanwise direction. The FTIR probe was aligned with the center injector, whereas the imaging measurement represents a line-of-sight average across the span. This effect is consistent with the larger differences observed for the discrete fueling cases.

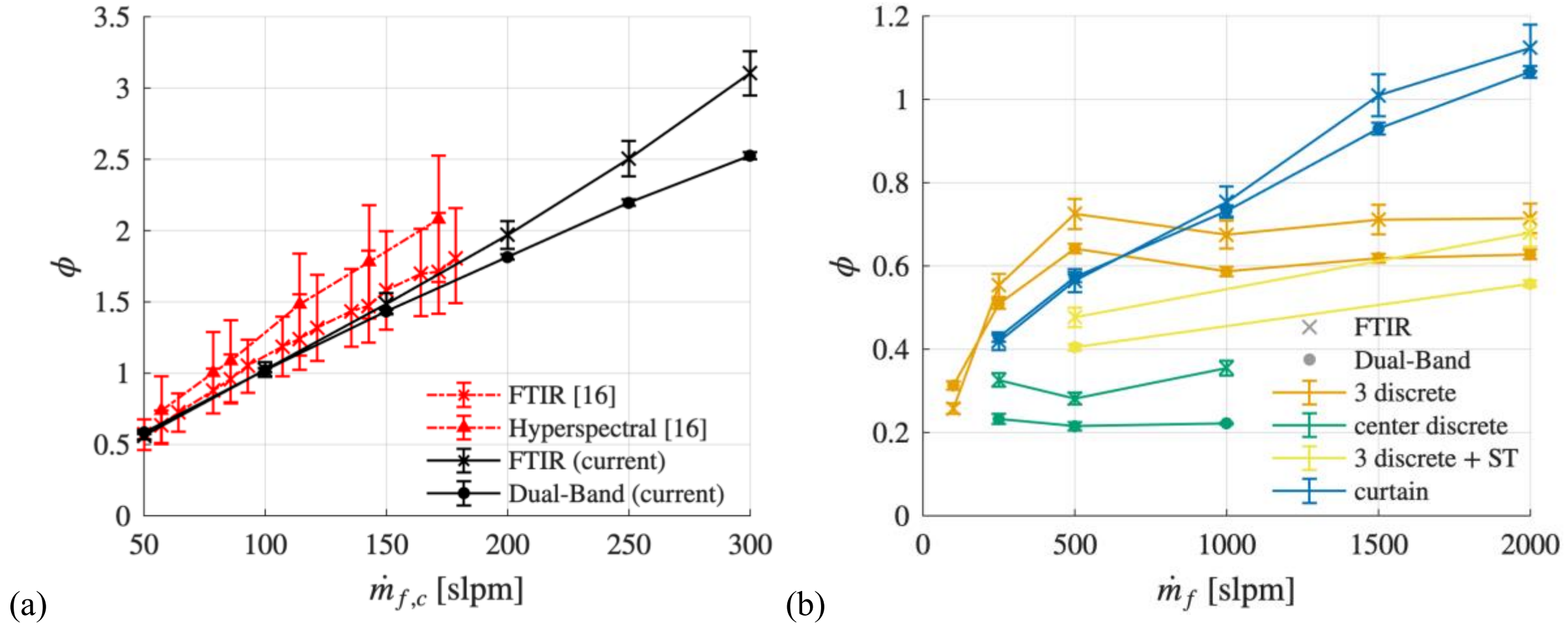

Figure 9: Equivalence ratio measurements within cavity at varied fuel rates comparing FTIR extractive sampling and the dual-band absorption technique: (a) cavity fueling, (b) remaining fueling configurations.

Figure 10 shows the measured temperature at the cavity and injector probe locations over the investigated fuel flow rate range. Previous hyperspectral measurements from Rhoby et al. [16] are included for the cavity fueling configuration. The hyperspectral measurements indicated little systematic change in cavity temperature with fuel flow rate because the variation remained within the measurement uncertainty. The present dual-band measurements show the same weak dependence on fuel flow rate but yield a cavity temperature near 500 K, compared with approximately 575 K for the previous hyperspectral measurements. The origin of this offset could not be isolated from the present data. Because the two methods agree much more closely at the injector location, the difference does not appear to represent a uniform temperature offset between the diagnostics.

At the cavity injector location, the previous hyperspectral measurements showed a monotonic decrease in temperature with increasing fuel flow rate. The dual-band measurements reproduce this trend and remain within approximately 10 K of the hyperspectral measurements over the investigated range. The cavity temperature also varied little with fuel flow rate for most of the other configurations. The principal exception was the center discrete injector, for which the cavity temperature decreased by approximately 50 K between 250 and 1000 slpm as the increasingly underexpanded jet introduced colder fuel into the flow.

The temperature at the injector locations showed a stronger dependence on fuel flow rate. For the three-discrete-injector configuration, the resolved temperature decreased from approximately 440 K to 370 K between 500 and 2000 slpm. The two lowest flow rate cases could not be resolved at the injector location because the fuel penetration and corresponding absorption signal were insufficient. The center discrete injector exhibited a similar trend and reached the lowest measured temperature, near 330 K. When compared at approximately matched flow rate per injector, the center and 3 discrete injector configurations gave nearly identical temperatures. For example, the 500 slpm center-injector case and the 1500 slpm three-injector case yielded temperatures of 376 and 378 K, respectively.

The ST cases exhibited a similar decrease in injector location temperature with increasing fuel flow rate but were shifted approximately 50 K higher than the corresponding cases without a ST. This increase indicates that compression associated with the shock train extended upstream to the injector region and reduced the degree of expansion cooling experienced by the injected fuel.

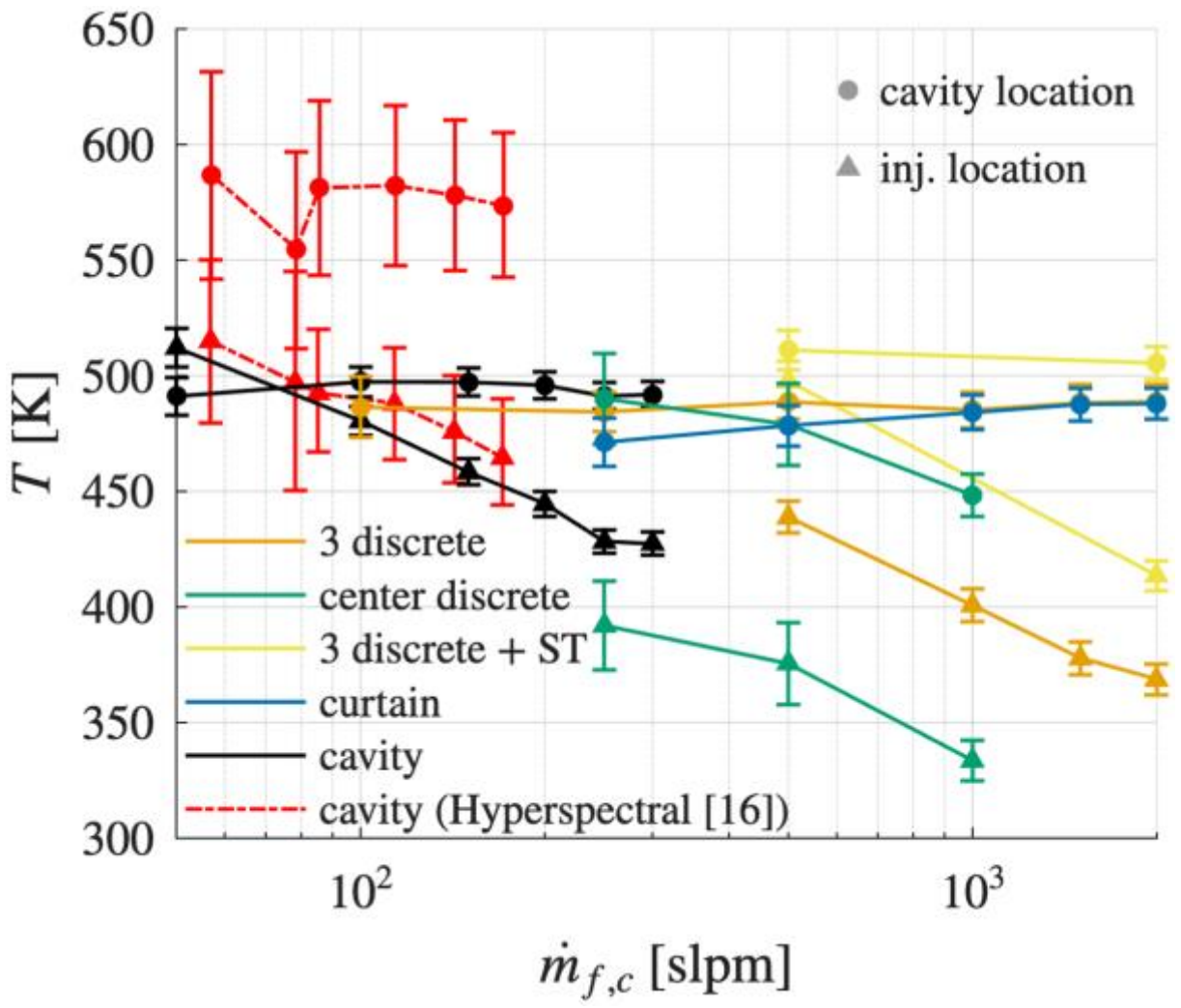

Figure 10: Temperature measured at the cavity and injector probe locations using the dual-band absorption technique. Previous hyperspectral measurements for cavity fueling from Rhoby et al. [16] are included for comparison.

The dual-band measurements show that fuel delivery to the cavity is governed by a competition between total fuel supply, jet penetration, and the surrounding flow structure. Curtain fueling produced a nearly proportional increase in cavity equivalence ratio with fuel flow rate, whereas discrete injection became progressively less effective at increasing cavity fueling as jet penetration increased. The shock train further redistributed the injected fuel while increasing the local gas temperature and density. The dual-band measurements captured these trends while providing spatially resolved concentration and temperature fields with substantially simpler spectral instrumentation than hyperspectral imaging.

### c. Unsteady Fuel Distributions

The dual-band measurement requires images from two spectral filters that are acquired sequentially rather than simultaneously. The concentration and temperature fields presented above were therefore determined from time-averaged images. For flows with significant temporal variation, instantaneous concentration fields can instead be obtained using a single spectral band for which the absorption cross-section is only weakly dependent on temperature. The 3385 nm filter satisfies this condition over the temperature range relevant to the present measurements. Its filter-weighted absorption cross-section was therefore approximated by a constant value of 1.99E-20 $cm^2$/molecule, and Equation (6) was solved directly for ethylene concentration for each acquired image. The same four-acquisition background correction was retained, with the time-averaged shielded fueled image used to remove the emission contribution from each instantaneous fueled image. The resulting fields were obtained at the 100 Hz filter-specific acquisition rate.

Figure 11 shows the temporal standard deviation of ethylene concentration normalized by its time-averaged value, $\sigma_C / \bar{C}$, for the three-discrete-injector configuration with and without a shock train at 500 slpm. Without the shock train, the cavity exhibited relatively little temporal variation, while larger fluctuations were concentrated along the fuel plume above the cavity. With the shock train present, the region of elevated fluctuation intensity broadened and followed the upward-deflected fuel plume. Increased variability was also observed within the cavity. The comparison indicates that unsteadiness associated with the shock train periodically changes the amount of fuel delivered to the cavity rather than simply altering its mean distribution.

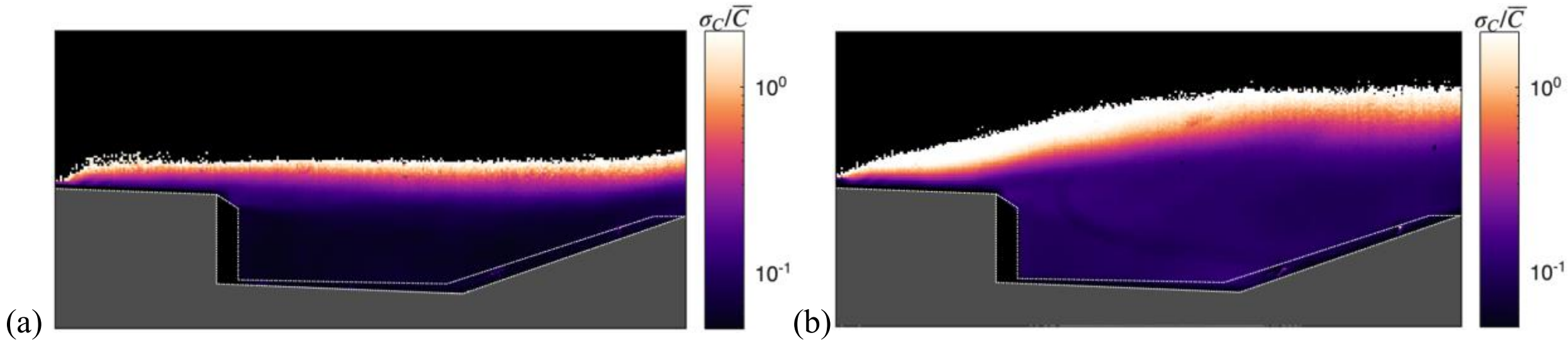

Figure 11: Normalized standard deviation of ethylene molar concentration at a fuel rate of 500 slpm comparing configurations (a) 3 discrete and (b) 3 discrete + ST.

The 3 discrete + ST case at 500 slpm was examined further in Fig. 12. Figure 12a shows instantaneous ethylene concentration fields at $t = 1.20$ and 1.44 s. At $t = 1.20$ s, the fuel remains close to the tunnel wall upstream of the cavity step and spreads downstream along the cavity shear layer. At $t = 1.44$ s, the plume is displaced farther from the wall and extends above the shear layer. The latter condition is accompanied by a visibly lower concentration within the cavity.

The temporal relationship between plume displacement and cavity fueling is shown in Fig. 12b. Intermittent increases in concentration at the probe above the upstream wall correspond to reductions in concentration at the cavity probe. The two example times in Fig. 12a illustrate these opposing states. The joint distribution of the two probe signals in Fig. 12c further shows a negative correlation between the concentrations at the two locations. Thus, upward displacement of the fuel plume by the unsteady STLE intermittently reduces the amount of fuel entering the cavity.

Such fluctuations can transiently shift the local cavity equivalence ratio even at fixed mean fuel flow rate and therefore may alter the conditions necessary for ignition and flameholding.

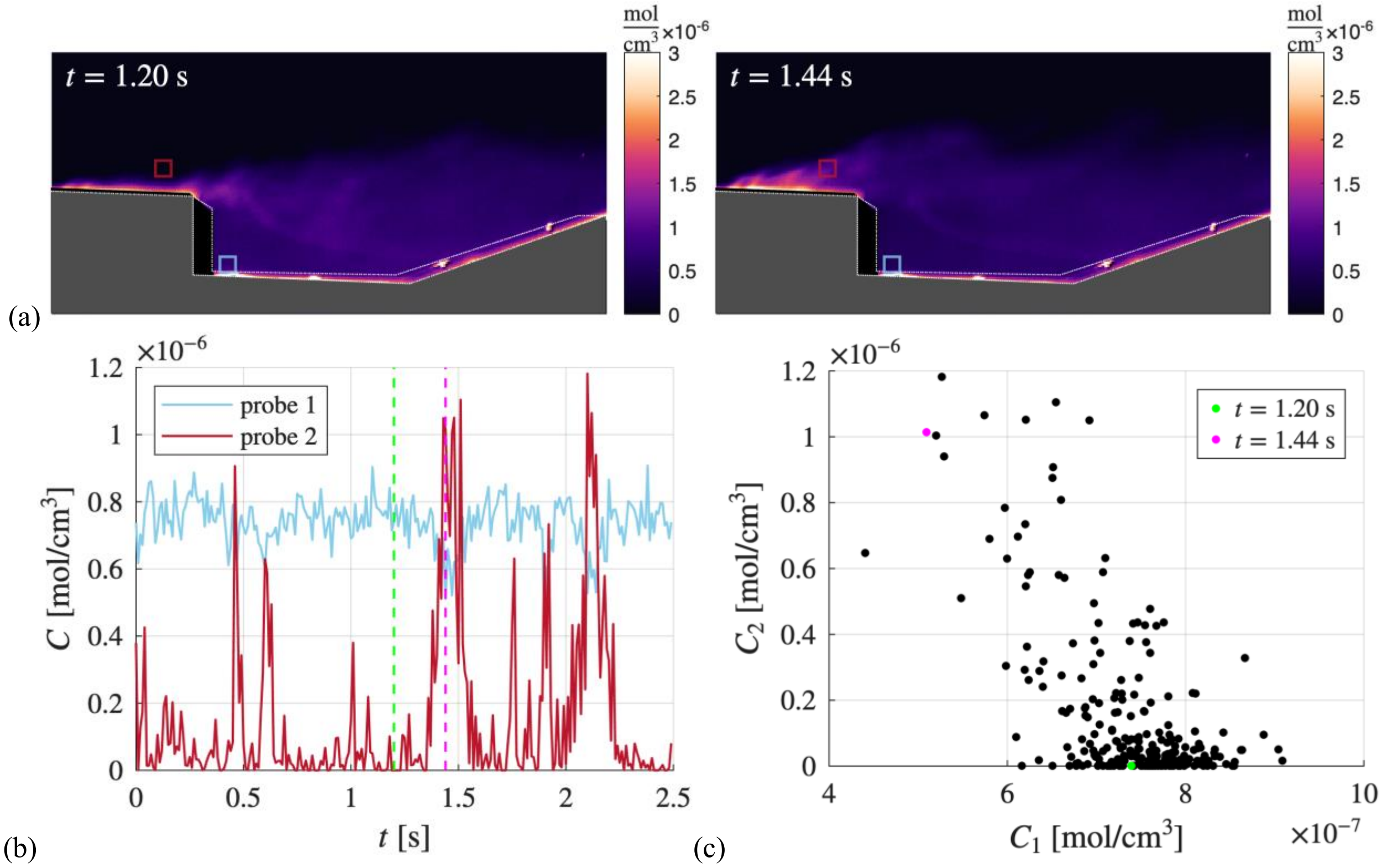

Figure 12: Unsteady analysis of the 3 discrete + ST configuration with a fuel rate of 500 slpm: (a) representative instantaneous molar concentration images with probe locations indicated, (b) molar concentration timeseries at probed locations, (c) phase portrait of molar concentration values at probed locations. Probed locations are denoted by the blue and red square markers. Time stamps are highlighted in pink and green.

## V. Conclusion

A dual-band mid-infrared absorption imaging method was developed for simultaneous two-dimensional measurement of hydrocarbon concentration and temperature. The method uses two narrowband filters with different temperature sensitivities to determine ethylene concentration and temperature from spatially resolved absorption measurements. Uncertainty associated with the temperature-dependent absorption cross-section fits and experimental image variability was propagated through the two-band inversion, and the 3365 and 3421 nm filter pair provided the lowest predicted temperature uncertainty over the range considered.

The method was demonstrated in a Mach-2 cavity flow using several ethylene injection configurations, and the resulting fields revealed distinct fueling behavior for each approach. Cavity injection produced high ethylene concentrations confined primarily within the cavity, with little penetration above the shear layer. Curtain injection introduced fuel broadly through the incoming boundary layer and produced a comparatively uniform cavity distribution whose equivalence ratio increased approximately linearly with fuel flow rate. Discrete injection produced greater penetration into the core flow; as injection rate increased, additional fuel increasingly passed above the cavity rather than being entrained into it, limiting the corresponding increase in cavity equivalence ratio. When a shock train was introduced, the shock train leading edge deflected the discrete fuel plume farther from the wall, reduced the cavity equivalence ratio, and increased the local temperature and density through compression. Across these configurations, the dual-band equivalence-ratio measurements agreed well with FTIR extractive measurements and reproduced the principal trends observed previously using hyperspectral imaging.

A single-band implementation was also demonstrated using the weakly temperature-dependent 3385 nm absorption band. Removing the requirement to combine two sequential filter measurements allowed instantaneous ethylene concentration fields to be obtained at the 100 Hz filter-specific acquisition rate, and could similarly be obtained at kHz-MHz rates depending on camera capability. Applied to the discrete-injection case with a shock train, these measurements showed that unsteadiness of the shock train leading edge intermittently displaced the fuel plume away from the wall. These plume deflections were correlated with decreases in cavity fuel concentration, demonstrating that the cavity equivalence ratio can fluctuate substantially even at fixed mean injection rate. The single-band implementation therefore provides a means of resolving transient fuel redistribution that is obscured by the time-averaged dual-band measurement. Figure 13 shows a diagram comparing the mid-IR absorption diagnostics discussed in this manuscript.

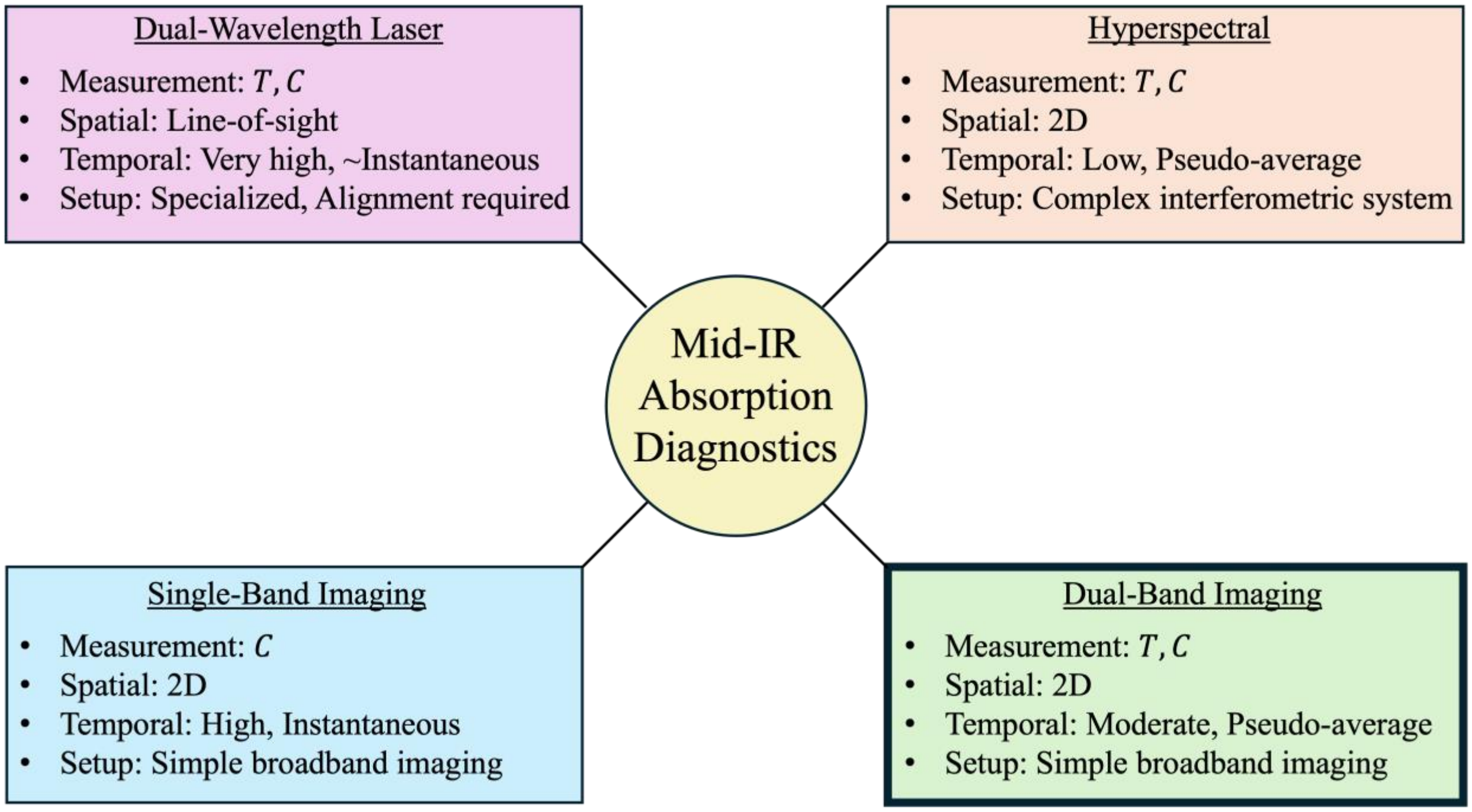

Figure 13: Mid-IR absorption diagnostics and their characteristics.

The dual-band approach occupies a useful middle ground between line-of-sight dual-wavelength absorption diagnostics and hyperspectral imaging. It retains quantitative, spatially resolved measurements of both species concentration and temperature while requiring only a broadband source, a mid-IR camera, and a small number of narrowband filters. The same optical system can also be operated in a single-band mode when temporal resolution is more important than simultaneous temperature measurement. This combination of quantitative accuracy, full-field spatial resolution, comparatively simple instrumentation, and flexible temporal capability makes narrowband mid-IR absorption imaging a practical diagnostic for complex compressible and reacting flows.

## Acknowledgments

The authors would like to thank Paul Gross and Scott Febo from Innovative Scientific Solutions, Inc. for assistance in operating the supersonic flow facility and execution of the experiments. This work was performed while the primary author held a National Research Council Research Associateship at the Air Force Research Laboratory.

## References

[1] Glassman, I., Yetter, R. A., & Glumac, N. G. (2014). *Combustion* (5th ed.). Academic Press.

[2] Ehn, A., Zhu, J., Li, X., & Kiefer, J. (2017). Advanced laser-based techniques for gas-phase diagnostics in combustion and aerospace engineering. *Applied Spectroscopy*, **71**(3), 341–366. https://doi.org/10.1177/0003702817690161

[3] Hanson, R. K. (1988). Combustion diagnostics: Planar imaging techniques. *Symposium (International) on Combustion*, **21**(1), 1677–1680. https://doi.org/10.1016/S0082-0784(88)80401-6

[4] Sick, V. (2013). High speed imaging in fundamental and applied combustion research. *Proceedings of the Combustion Institute*, **34**(2), 3509–3530. https://doi.org/10.1016/j.proci.2012.08.012

[5] Lackner, M. (2007). Tunable diode laser absorption spectroscopy (TDLAS) in the process industries–a review. *Reviews in Chemical Engineering*, **23**(2), 65–147.

[6] Donbar, J. M., Driscoll, J. F., & Carter, C. D. (2000). Reaction zone structure in turbulent nonpremixed jet flames—From CH-OH PLIF images. *Combustion and Flame*, **122**(1–2), 1–19. https://doi.org/10.1016/S0010-2180(00)00098-5

[7] Do, H., & Carter, C. D. (2013). Hydrocarbon fuel concentration measurement in reacting flows using short-gated emission spectra of laser induced plasma. *Combustion and Flame*, **160**(3), 601–609. https://doi.org/10.1016/j.combustflame.2012.12.002

[8] McGann, B., Ombrello, T. M., Peterson, D. M., Hassan, E., Hammack, S. D., Carter, C. D., Lee, T., & Do, H. (2021). Lean fuel detection with nanosecond-gated laser-induced breakdown spectroscopy. *Combustion and Flame*, **224**, 209–218. https://doi.org/10.1016/j.combustflame.2020.12.014

[9] Miller, J. D., Slipchenko, M., Meyer, T. R., Jiang, N., Lempert, W. R., & Gord, J. R. (2009). Ultrahigh-frame-rate OH fluorescence imaging in turbulent flames using a burst-mode optical parametric oscillator. *Optics Letters*, **34**(9), 1309. https://doi.org/10.1364/OL.34.001309

[10] Blunck, D., Basu, S., Zheng, Y., Katta, V., & Gore, J. (2009). Simultaneous water vapor concentration and temperature measurements in unsteady hydrogen flames. *Proceedings of the Combustion Institute*, **32**(2), 2527–2534. https://doi.org/10.1016/j.proci.2008.05.046

[11] Rankin, B. A., Ihme, M., & Gore, J. P. (2015). Quantitative model-based imaging of mid-infrared radiation from a turbulent nonpremixed jet flame and plume. *Combustion and Flame*, **162**(4), 1275–1283. https://doi.org/10.1016/j.combustflame.2014.10.019

[12] Gross, K. C., Bradley, K. C., & Perram, G. P. (2010). Remote identification and quantification of industrial smokestack effluents via imaging Fourier-transform spectroscopy. *Environmental Science & Technology*, **44**(24), 9390–9397. https://doi.org/10.1021/es101823z

[13] Ombrello, T., Blunck, D. L., & Resor, M. (2016). Quantified infrared imaging of ignition and combustion in a supersonic flow. *Experiments in Fluids*, **57**(9), 140. https://doi.org/10.1007/s00348-016-2210-0

[14] Childs, N. A., Ombrello, T., & Wolff, M. (2025, January 6). Mid-Infrared Imaging of Supersonic Combustion Exhaust. *AIAA SCITECH 2025 Forum*. https://doi.org/10.2514/6.2025-0227

[15] De Donato, P., Barres, O., Sausse, J., & Martin, D. (2018). Near real-time ground-to-ground infrared remote-sensing combination and inexpensive visible camera observations applied to tomographic stack emission measurements. *Remote Sensing*, **10**(5), 678. https://doi.org/10.3390/rs10050678

[16] Rhoby, M. R., Ombrello, T. M., Okhovat, N. S., Hansen, A. M., & Gross, K. C. (2018). Infrared Hyperspectral Imaging Diagnostics of Fueling Strategy for Scramjets. *Journal of Propulsion and Power*, **34**(6), 1391–1400. https://doi.org/10.2514/1.B36939

[17] Ombrello, T., Okhovat, N. S., and Rhoby, M. R. (2018). "Measurements of Scramjet Fueling Conditions," *2018 AIAA Aerospace Sciences Meeting*, AIAA Paper 2018-1360, American Institute of Aeronautics and Astronautics, Kissimmee, FL, Jan. 2018. https://doi.org/10.2514/6.2018-1360

[18] van der Lee, J., Senior-Tybora, W., Kaner, R., Aldema-Tshuva, Y., Michaels, D., and Lefkowitz, J. K. (2026). "Mixing and Entrainment of Liquid Fuel in a Cavity-Based Scramjet Model Combustor," *Experimental Thermal and Fluid Science*, **172**, Art. No. 111629. https://doi.org/10.1016/j.expthermflusci.2025.111629

[19] Klingbeil, A. E., Porter, J. M., Jeffries, J. B., & Hanson, R. K. (2009). Two-wavelength mid-IR absorption diagnostic for simultaneous measurement of temperature and hydrocarbon fuel concentration. *Proceedings of the Combustion Institute*, **32**(1), 821–829.

[20] Pyun, S. H., Porter, J. M., Jeffries, J. B., Hanson, R. K., Montoya, J. C., Allen, M. G., & Sholes, K. R. (2009). Two-color-absorption sensor for time-resolved measurements of gasoline concentration and temperature. *Applied optics*, *48*(33), 6492-6500.

[21] Gruber, M. R., & Nejad, A. S. (1995). New supersonic combustion research facility. *Journal of propulsion and power*, *11*(5), 1080-1083.

[22] Alrefae, M., Es-sebbar, E. T., & Farooq, A. (2014). Absorption cross-section measurements of methane, ethane, ethylene and methanol at high temperatures. *Journal of Molecular Spectroscopy*, *303*, 8-14.

[23] Matsuo, K., Miyazato, Y., & Kim, H. D. (1999). Shock train and pseudo-shock phenomena in internal gas flows. *Progress in aerospace sciences*, *35*(1), 33-100.